\documentclass[twocolumn,showpacs,preprintnumbers,amsmath,amssymb]{revtex4}
\usepackage{graphicx}
\usepackage{dcolumn}
\usepackage{bm}

\begin{document}

\preprint{}

\title{Deterministic single-atom excitation via adiabatic passage and 
Rydberg blockade}
\author{I.~I.~Beterov}
\email{beterov@isp.nsc.ru}
\author{D.~B.~Tretyakov}
\author{V.~M.~Entin}
\author{E.~A.~Yakshina}
\author{I.~I.~Ryabtsev}
  
\affiliation{A.V.Rzhanov Institute of Semiconductor Physics SB RAS, Prospekt Lavrentieva 13, 630090 Novosibirsk, Russia}
\author{C.~MacCormick}
\author{S.~Bergamini}
\affiliation{The Open University, Walton Hall, MK6 7AA, Milton Keynes, UK}

\date{25 July 2011}

\begin{abstract}
We propose to use adiabatic rapid passage with a chirped laser pulse in the 
strong dipole blockade regime to deterministically excite only one Rydberg 
atom  from randomly loaded optical dipole traps or optical lattices. The 
chirped laser excitation is shown to be insensitive to the random number 
\textit{N} of the atoms in the traps. Our method overcomes the problem of the $\sqrt {N} $ 
dependence of the collective Rabi frequency, which was the main obstacle for 
deterministic single-atom excitation in the ensembles with unknown \textit{N}, and can be applied for single-atom loading of 
dipole traps and optical lattices.
\end{abstract}


\maketitle

\section{Introduction}

Dipole blockade of the laser excitation of neutral atoms to highly excited Rydberg states~\cite{Lukin} opens up new opportunities for entanglement 
engineering and quantum information processing~\cite{SaffmanRev}. In mesoscopic ensembles of strongly interacting 
neutral atoms the dipole blockade manifests itself as a suppression of the excitation of more than one atom by narrow-band laser radiation, due to the 
shifts of the Rydberg energy levels induced by long-range interactions~\cite{Lukin}. Recently, dipole blockade was observed for two Rb atoms trapped in two 
spatially separated optical dipole traps~\cite{Wilk,SaffmanCNOT,UrbanNature}. 

The deterministic excitation of a single atom from a mesoscopic ensemble into a Rydberg state using dipole blockade can be exploited in a variety of applciations, as the creation of quantum information processor for photons~\cite{Honer} or single-atom loading of  optical lattices or optical dipole traps~\cite{Saffman2002}, which is of crucial importance for the development of quantum registers~\cite{SaffmanRev}, single-photon sources~\cite{Photon}, high-precision metrology in optical lattice clocks~\cite{Clock}, and phase transitions in artificial solid structures with Rydberg excitations~\cite{Solid}. 

Single-atom loading remains a challenge since no simple and reliable method for the preparation of the optical lattice with single occupancy at each site is available yet. So far, only the Mott insulator regime in Bose-Einstein condensates (BEC) has demonstrated its ability to provide single-atom loading of large-scale optical lattices~\cite{Mott}. However, obtaining a BEC is a complicated and slow procedure, which may be not well suited for fast quantum computation. Another approach for single-atom loading of multiple sites is the exploitation of a collisional blockade mechanism~\cite{ColBlock},  but it suffers from the  low loading efficiency for large arrays. Recently, the fidelity of single atom loading of $82.7\%$ was demonstrated using light-assisted collisions~\cite{ColBlock2}. Although being an important step forward, this fidelity is not enough for the creation of a scalable quantum register~\cite{SaffmanRev}.

Highly excited Rydberg atoms with the principal quantum number $n>>1$~\cite{Gallagher} can be used to implement fast quantum logic gates~\cite{Jaksch, Lukin}. These atoms exhibit strong dipole-dipole interaction at distances that can be as large as a few microns. Therefore, dipole-dipole interaction can also be used in schemes for single-atom loading of optical lattices and traps arrays, since the interatomic spacing in lattices sites lies in the micron range.

The first proposal for single-atom loading exploiting a dipole blockade of the laser excitation of mesoscopic ensembles of $N$ cold ground-state atoms~\cite{Lukin, Comparat} was formulated in Ref.~\cite{Saffman2002}.
A strong dipole blockade was suggested to provide deterministically a single Rydberg atom, while the remaining $N-1$ ground-state atoms could be selectively removed from the lattice site by an additional laser pulse.
In Ref.~\cite{SaffmanRev} it has been pointed out, however, that this method demands identical initial numbers of cold atoms in each lattice site, because collective Rabi frequency  of  single-atom excitation  depends on $N$:

\begin{equation}
\Omega_N = \Omega_1 \sqrt{N}
\end{equation}

\noindent(here $\Omega_1$ is the Rabi frequency for a single atom). This requirement is not fulfilled in optical lattices, which are loaded from cold atom clouds at random and typically have a Poissonian distribution of the number of atoms in each site \cite{Poisson}.

Full dipole blockade ensures that an ensemble of \textit{N} atoms shares a collective single excitation oscillating between the ground and Rydberg state at the Rabi frequency $\Omega_N$~\cite{Gaetan}. For a given number \textit{N} of atoms 
experiencing full dipole blockade, it is possible to excite a single atom 
into the Rydberg state using a monochromatic laser $\pi$-pulse of duration 
\begin{equation}
\tau_N=\pi/\left(\Omega_1\sqrt{N}\right).  
\end{equation}

\noindent However, in randomly loaded traps or optical lattices the atom-number uncertainty for 
the Poissonian statistics is $\Delta N \approx \sqrt {\overline {N}} $, where 
$\overline {N}$ is the mean number of atoms in the traps. Therefore, although the 
suppression of the excitation of more than one atom in the trap is 
guaranteed in the full dipole blockade regime, the value of $\tau_{N} $ 
required for the deterministic single-atom excitation is uncertain due to 
the uncertainty of \textit{N} in the individual traps~\cite{SaffmanRev}.

Adiabatic passage by sweeping the laser frequency through the resonance~\cite{Loy} or by counter-intuitive sequence of two monochromatic pulses known as Stimulated Raman Adiabatic Passage or STIRAP~\cite{Bergmann} is widely used to obtain a population inversion in multi-level systems due to their insensitivity to the Rabi frequencies of the particular transitions, provided the adiabaticity condition is satisfied~\cite{Bergmann,ChirpTheory}. We therefore may expect that adiabatic passage is also suitable to the blockaded ensembles with unknown number of atoms. However, recently it has been found that STIRAP with zero detuning from the intermediate state does not provide deterministic single-atom excitation in a blockaded ensemble~\cite{Moller}. In this work we propose to deterministically excite a single Rydberg atom using a chirped laser pulse or off-resonant STIRAP. 

\section{Deterministic single-atom excitation with chirped laser pulses}

 The time dependence of the amplitude and frequency for the chirped laser pulse is shown in Fig.~1(a). The laser frequency is linearly swept across the resonance during the pulse. In contrast to the results of Ref.~\cite{Moller}, we have found that chirped laser excitation  deterministically transfers the population of the blockaded ensemble into the collective state, which shares a single Rydberg excitation, as shown in Fig.1~(b).

\begin{figure}
\includegraphics[scale=0.5]{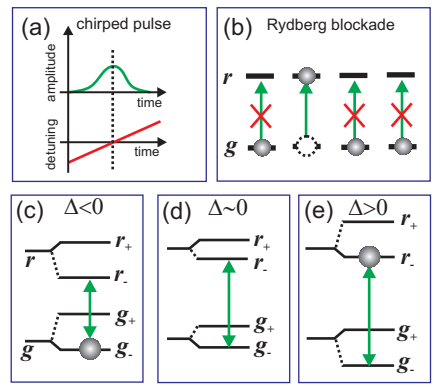}
\caption{
\label{Fig1}(Color online). 
(a) The time dependence of the amplitude and detuning 
from the resonance with atomic transition for a chirped pulse; (b) Scheme of 
the deterministic single-atom excitation from ground state $g$ to Rydberg 
state $r$ in a blockaded Rydberg ensemble. (c),(d),(e) Scheme of the 
adiabatic rapid passage. Energies of the dressed states are shown for laser 
detunings (c) $\Delta< 0$, (d) $\Delta \sim 0$, and (e) $\Delta > 0$. Laser 
frequency is rapidly chirped across the resonance. The population is 
transferred from state $g_{-}  $ to state $r_{-}  $ independently of the 
Rabi frequency.}
\end{figure}

The principle of chirped excitation can be understood as follows~\cite{Shalagin}. The dressed state 
energies for a two-level atom are shown in Figs.~1(c)-1(e) for laser detunings $\Delta < 0$, 
$\Delta \sim 0$ and $\Delta > 0$, respectively. At large negative detuning 
$\Delta < 0$ the energy of the unperturbed ground state \textit{g} is close 
to the energy of the dressed state $g_{-}  $, whereas at large positive 
detuning $\Delta > 0$ the energy of the dressed state $r_{-}  $ lies close 
to the energy of the unperturbed Rydberg state $r$. Initially, only the ground  $g_{-}$ 
state is populated [Fig.~1(c)]. When the laser frequency is swept across 
the resonance [Fig.~1(d)] the system adiabatically follows the dressed $\left\{-\right\}$ state 
of the dressed system, and therefore populates the Rydberg state $r_{-}$ after the 
end of the chirped laser pulse [Fig.~1(e)].

In order to show that chirped 
laser excitation in the blockaded ensemble is insensitive to \textit{N}, we 
have performed numerical simulations of the dipole blockade for $37P_{3/2}$
Rydberg state in Rb atoms. Rydberg atoms in an identical state \textit{nL} 
may interact via a F$\ddot{\mathrm{o}}$rster resonance if this state lies midway between two 
other levels of the opposite parity~\cite{Vogt}. The $37P_{3/2}$ state has a 
convenient Stark-tuned F$\ddot{\mathrm{o}}$rster resonance $37P_{3/2} + 37P_{3/2} \to 37S_{1/2} + 38S_{1/2}$, which we 
studied earlier in detail \cite{RyabtsevPRL,MonteCarlo}. 

We consider an excitation of the $37P_{3/2}$ state by a linearly-chirped 
Gaussian laser pulse (Fig.~2). In the time domain its electric field is expressed
as
\begin{equation}
\label{eq1}
E\left({t} \right) = E_{0} \mathrm{exp}\left[ {\frac{{ - t^{2}}}{{2\tau ^{2}}}} 
\right]\mathrm{cos}\left[ {\omega _{0} t + \alpha \frac{{t^{2}}}{{2}}} \right].
\end{equation}
\noindent Here $E_0$ is the peak electric field at $t = 0$, $\omega_{0}$ is the frequency of the atomic transition, $\tau = 1\,\mu \mathrm{s}$ is 
the half-width at $1/e$ intensity [Fig.~2(a)], and $\alpha $ is the 
chirp rate~\cite{ChirpTheory}. We choose $E_{0} $ to be such as to provide a single-atom 
peak Rabi frequency $\Omega_1 /\left( {2\pi}  \right) = 2$~MHz or $\Omega_1 
/\left( {2\pi}  \right) = 0.5$~MHz at the $5S \to 37P_{3/2}$ optical 
transition in Rb atoms. For convenience, the central frequency of the laser 
pulse is taken to be exactly resonant with the atomic transition at the 
maximum of the pulse amplitude. The atoms begin to interact with the laser 
radiation at $t = - 4\;\mu \mathrm{s}$.

The adiabaticity condition for a chirped pulse exciting a single Rydberg 
atom is given by~\cite{ChirpTheory}:
\begin{equation}
\label{eq2}
\left|d\Delta/dt\right|\ll\Omega_1^2
\end{equation}
\noindent For $N > 1$ the collective Rabi frequency $\Omega_N = \Omega_1\sqrt{N}$
grows with \textit{N}. Hence, we must only fulfill the adiabaticity 
condition for the excitation of a single atom.

\begin{figure}
\includegraphics[scale=0.4]{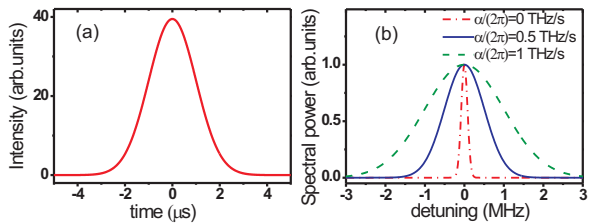}
\caption{\label{Fig2}
(Color online). (a) Envelope and (b) spectrum of the chirped Gaussian laser pulses with the chirp rates  $\alpha/\left(2\pi\right)=0.5$~THz/s (solid line) and  $\alpha/\left(2\pi\right)=1$~THz/s  (dashed line) used in numerical calculations.  The spectrum of 
the unchirped laser pulse of the same duration is shown for reference 
(dash-dotted line, height rescaled for clarity).
}
\end{figure}

The envelope of the laser pulse is a Gaussian that ensures the adiabatic switching of the laser-atom 
interaction. Figure~2(b) shows the calculated spectra of the laser pulses with $\alpha/\left(2\pi\right)=0.5$~THz/s  
(solid line) and  $\alpha/\left(2\pi\right)=1$~THz/s    (dashed line). The spectrum is broadened to the FWHM of  
$\Delta\omega/\left(2\pi\right)=1.2$~MHz    at  $\alpha/\left(2\pi\right)=0.5$~THz/s and $\Delta\omega/\left(2\pi\right)=2.4$~MHz  at $\alpha/\left(2\pi\right)=1$~THz/s
due to the frequency chirp, as can be seen from the comparison with the unchirped Gaussian pulse 
[$\Delta\omega/\left(2\pi\right)=0.2$~MHz] shown as the dash-dotted line in Fig.~2(b). This broadening could affect 
the dipole blockade efficiency and lead to leakage of the population 
to the collective states with more than one excitation. We therefore have 
performed a numerical calculation of the blockade efficiency for chirped 
laser excitation of an ensemble consisting of \textit{N}=1-7 atoms. The time-dependent Schr$\ddot{\mathrm{o}}$dinger equation was solved for the amplitudes of the 
collective states, taking into account all possible binary interactions 
between Rydberg atoms~\cite{RyabtsevPRL,MonteCarlo}. 

The calculations have been done for the exact 
Stark-tuned F$\ddot{\mathrm{o}}$rster resonance $37P_{3/2} + 37P_{3/2} \to 37S_{1/2} + 38S_{1/2}$ with zero energy 
defect which is supposed to be tuned by the electric field of 1.79~V/cm~\cite{RyabtsevPRL}. The numerically calculated time dependencies of the probability $P_1$ to 
excite a single Rydberg atom by the chirped laser pulse are shown in Fig.~3 
for $\alpha/\left(2\pi\right)=1$~THz/s, $\Omega_1/\left(2 \pi \right)=2$~MHz  (the left-hand 
panels) and for $\alpha/\left(2\pi\right)=0.5$~THz/s, $\Omega_1/\left(2 \pi \right)=0.5$~MHz  (the right-hand 
panels). Figures~3(a)-3(b) correspond to \textit{N}=1 (a single non-interacting atom) 
and serve as the references to compare with the interacting atoms. The 
transition probability in Fig.~3(a) is nearly unity with accuracy better 
than 0.02\%, while in Fig.~3(b) it reaches 0.993 at the end of the laser 
pulse. However, below we will show that for single-atom excitation at 
$N>1$ the conditions of Fig.~3(b) are preferred.

\begin{figure}
\includegraphics[scale=0.4]{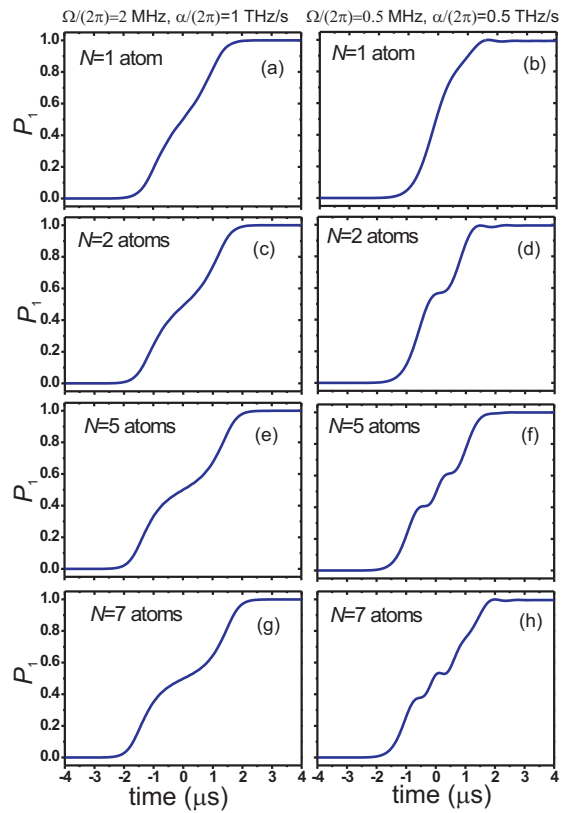}
\caption{
\label{Fig3}(Color online). 
Time dependence of the probability $P_1$ 
to excite a single Rb($37P_{3/2}$) atom in a trap containing 
(a)-(b)\textit{} 1~atom, (c)-(d) 2~atoms, (e)-(f) 5~atoms, and (g)-(h) 7~atoms by chirped Gaussian laser pulses. In the left-hand panels the chirp 
rate is  $\alpha/\left(2\pi\right)=1$~THz/s and Rabi frequency is $\Omega_1/\left(2 \pi \right)=2$~MHz. In the right-hand   panels the chirp rate is  $\alpha/\left(2\pi\right)=0.5$~THz/s and Rabi frequency $\Omega_1/\left(2 \pi \right)=0.5$~MHz.
The calculations have been done for the exact Stark-tuned F$\ddot{\mathrm{o}}$rster resonance $37P_{3/2} + 37P_{3/2} \to 37S_{1/2} + 
38S_{1/2}$ with zero energy defect. The atoms are randomly placed in the cubic 
interaction volume of the size $L = 1\;\mu \mathrm{m}$.}
\end{figure}


The calculated time dependencies of $P_1$ in the full dipole blockade regime are shown 
in Fig.~3 for 2~atoms [(c)-(d)], 5~atoms [(e)-(f)], and 7~atoms [(g)-(h)]. The 
\textit{N} atoms were randomly located in a $L \times L \times L\;{\kern 
1pt} \mu \mathrm{m}^{3}$ cubic volume with $L = 1\;\mu \mathrm{m}$. The full blockade regime 
was evidenced by complete suppression of the probability to excite more than 
one atom. The calculations have shown that the fidelity $P_1$ of the 
population inversion at $t = 4\;\mu \mathrm{s}$ reaches 99\% regardless of 
\textit{N}.  This is the main result of this article that confirms that our 
proposal can be implemented in practice.

Surprisingly, the effect of chirped laser excitation in the blockaded ensemble is completely different from the effect of on-resonant STIRAP, discussed in Ref.~\cite{Moller}. The scheme of the three-level ladder system interacting with the two laser fields with Rabi frequencies $\Omega_s$ and $\Omega_c$ is shown in Fig.~4(a). We consider an exact two-photon resonance for the ground state $g$ and Rydberg state $r$, while the intermediate excited state $e$ has in general a variable detuning $\delta$ from the resonance both with $\Omega_s$ and $\Omega_c$ fields (but with the opposite signs). The time sequence of the two Gaussian pulses with halfwidths $\tau_c$ and $\tau_s$ at $1/e$ intensity and time interval $\Delta T$ is shown in Fig.~4(b). For simplicity, we have chosen identical Rabi frequencies $\Omega_s/\left(2\pi\right)=\Omega_c/\left(2\pi\right)=10$~MHz. The pulse halfwidths $\tau_c=\tau_s=1$~$\mathrm{\mu s}$ were also identical to the half of the interval between the pulses $\Delta T=2$~$\mathrm{\mu s}$~\cite{Bergmann}. 

We have calculated the excitation probabilities of the collective states of an ensemble consisting of two interacting atoms in a full blockade regime. The time dependencies of the probability of single-atom Rydberg excitation $P_1$ (solid line) and the probabilities  $P_{ee}$ (both atoms in the lower excited state, dashed line), $P_{gg}$ (both atoms in the ground state, dash-dotted line) are shown for on-resonant STIRAP with $\delta=0$ in Fig.~4(c) and for off-resonant STIRAP with detuning from the intermediate state $\delta=10$~MHz~in Fig.~4(d). Figure~4(c) qualitatively reproduces the results of Ref.~\cite{Moller} for two atoms. After the end of the adiabatic passage the population is redistributed between $g$ and $e$ states, while Rydberg state $r$ remains unpopulated. We have found that the observed breakdown of STIRAP in the blockaded ensemble results from the destructive interference of laser-induced transitions in the quasimolecule consisting of interacting atoms, and it can be avoided by an increase of the detuning from the intermediate state, which finally makes STIRAP equivalent to chirped excitation~\cite{Shore}. Figure~4(d) shows that as well as chirped excitation, off-resonant STIRAP can be used for deterministic excitation of a single Rydberg atom.

\begin{figure}
\includegraphics[scale=0.45]{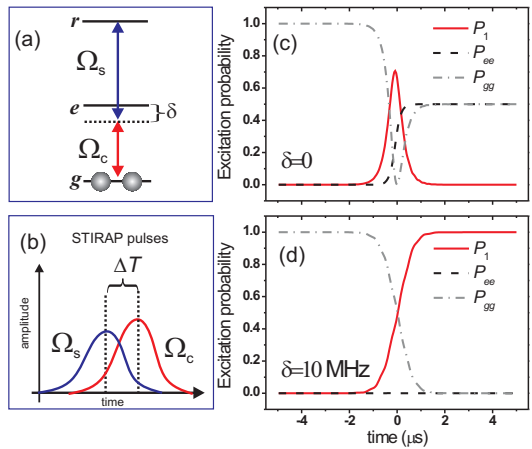}
\caption{
\label{Fig4}(Color online). 
(a) Scheme of the three-level ladder system interacting with the two laser fields with Rabi frequencies $\Omega_s$ and $\Omega_c$ and detuning from the intermediate state $\delta$. (b) Time sequence of pulses for STIRAP. The delay between the two pulses is $\Delta T$. (c) Time dependencies of the probabilities $P_{gg}$,  $P_{ee}$ and the probability to excite a single Rydberg atom $P_1$ for on-resonant STIRAP with $\delta=0$. (d) Time dependencies of the probabilities $P_{gg}$,  $P_{ee}$ and the probability to excite a single Rydberg atom $P_1$ for off-resonant STIRAP with $\delta=10$~MHz.}
\end{figure}


\section{Fidelity of deterministic single-atom excitation}

The main limitation of the proposed method is a possible breakdown of the 
full dipole blockade in the realistic experimental conditions. The 
\textit{N} atoms in an optical dipole trap have a finite temperature and are 
located randomly due to atomic motion. The blockade breakdown for two 
interacting Rydberg atoms can be caused by the weakness of dipole-dipole 
interaction between remote atoms or by more complicated mechanisms, 
including zeros of F$\ddot{\mathrm{o}}$rster resonances~\cite{ForsterZeros} and destructive interference in 
many-atom ensembles~\cite{PohlBerman}.

The fidelity of the single-atom excitation can be defined as the probability 
$P_1$ to have exactly one atom excited at the end of the laser pulse. We 
have numerically calculated $P_1$ for various sizes of the atomic sample 
in an optical dipole trap for chirped laser excitation. The \textit{N} atoms were randomly located in a 
$L \times L \times L\;{\kern 1pt} \mu \mathrm{m}^3$ cubic volume. The value of 
$P_1$ was averaged over $\sim 10^4$ random spatial configurations.

The dependencies of $P_1$ on \textit{L} at the end of the laser pulse ($t= 4\;\mu \mathrm{s}$) are shown in Fig.~5(a) for 
$\alpha/\left(2\pi\right)=1$~THz/s, $\Omega_1/\left(2 \pi \right) = 2$~MHz and 2~(squares), 3~(circles), 4~(triangles), and 
5~(rhombs) atoms. For $L > 1\;\mu \mathrm{m}$ we have found that $P_1$ reduces as 
\textit{L} increases, mostly due to the fluctuations of the spatial 
positions of the atoms in a disordered sample~\cite{MonteCarlo}. More surprisingly, we 
have found that $P_1$ depends on \textit{N} in a counter-intuitive way: 
it drops as \textit{N} increases. This observation is 
presumably due to the quantum interference between different energy exchange 
channels in many-atom ensembles, which was discussed in Ref.~\cite{PohlBerman}. Our Monte Carlo numerical simulations in Fig.5~(a) and (b), which accounted also for the F$\ddot{\mathrm{o}}$rster zeros~\cite{ForsterZeros}, have shown that they do not affect the dipole blockade even for atoms randomly placed within the interaction volume, if the Rydberg interactions are strong enough, i.e., when the blockade shift is larger than the laser linewidth.

The same calculations have been done for $\alpha/\left(2\pi\right)=0.5$~THz/s and $\Omega_1/\left(2 \pi \right)=0.5$~MHz [Fig.~5(b)]. Although in these conditions 
$P_1$ also reduces with the increase of \textit{L} and \textit{N}, the 
efficiency of the dipole blockade is better, than in Fig.~5(a). For all 
\textit{N} we have found that in Fig.~5(b) the fidelity $P_1\ge 0.95$ is 
achieved at $L \le 2\;\mu \mathrm{m}$, while in Fig.~5(a) a localization $L \le 1\;\mu \mathrm{m}$ is required. This difference can be crucial 
in the real experimental conditions, since it is difficult to localize the 
atoms in the volumes of the size comparable with optical wavelengths. 

The increase of the chirp rate and Rabi frequency may be desirable to reduce 
the excitation time and to avoid the errors due to finite lifetimes of 
Rydberg states~\cite{Lifetimes}. We have found, however, that such an increase would 
decrease the blockade efficiency, which depends on both $\Omega_1$ and 
$\alpha$. To find the optimal values of $\Omega_1$ and $\alpha$, we 
have first calculated the dependencies of $P_1$ on $\Omega_1$ and 
$\alpha$ for the two frozen atoms in the full blockade regime, which can be 
modeled simply by increasing the effective Rabi frequency $\Omega_2= 
\Omega_1 \sqrt {2}$ in a single two-level atom. This dependence is 
presented in Fig.~5(c) as a density plot, as in Ref.~\cite{ChirpTheory}. The light areas in 
Fig.~5(c) show the regions where $P_1\approx 1$. The periodic structure 
across the $\alpha = 0$ axis represents the coherent Rabi oscillations at 
the $n\pi $ laser pulses with zero chirp. The area of the robust rapid 
adiabatic passage at $\alpha \ll 1/\tau^2$ is limited by 
the adiabaticity condition of Eq.~(\ref{eq2}). However, a possible breakdown of the 
dipole blockade adds more restrictions on the values of $\Omega_1$ and 
$\alpha$. The dependence of $P_1$ on $\Omega_1$ and 
$\alpha$ for the two interacting atoms randomly placed in the cubic volume 
with $L = 4\;\mu \mathrm{m}$ is shown in Fig.~5(d). The probability $P_1$ drops 
with the increase of both $\Omega_1$ and $\alpha$. However, it remains 
nearly constant in the region between the coherent and adiabatic regimes 
with small chirp rate $0.3$~THz/s$\le \left|\alpha/\left(2\pi\right)\right|\le 0.7$~THz/s.
For our F$\ddot{\mathrm{o}}$rster resonance $37P_{3/2} + 37P_{3/2} \to 37S_{1/2} + 38S_{1/2}$ and the pulse 
width $\tau = 1\,\mu \mathrm{s}$ we have also found the optimal Rabi frequency to be 
$0.4$~MHz$\le\Omega_1/\left(2 \pi \right) \le 0.6$~MHz. 
\begin{figure}
\includegraphics[scale=0.4]{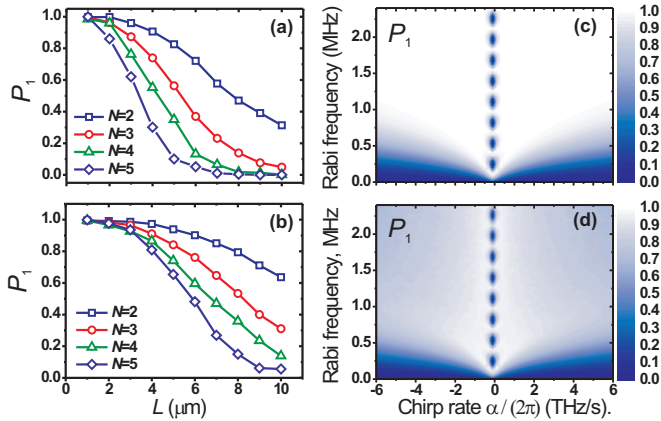}
\caption{\label{Fig5}(Color online). (a) Dependencies of the probability $P_{1} $ to 
excite a single Rb($37P_{3/2}$) atom on the size \textit{L} of the cubic 
interaction volume for 2~(squares), 3~(circles), 4~(triangles), and 5~(rhombs) randomly positioned atoms at the chirp rate $\alpha/\left(2\pi\right)=1$~THz/s and Rabi frequency $\Omega_1/\left(2 \pi \right)=2$~MHz. (b) The same dependencies at the chirp rate $\alpha/\left(2\pi\right)=0.5$~THz/s and Rabi frequency $\Omega_1/\left(2 \pi \right)=0.5$~MHz.
(c) Dependence of $P_1$ on Rabi frequency and chirp rate for the two 
frozen atoms in the full blockade regime. (d) Monte-Carlo simulation of the 
same dependence for the two interacting atoms, randomly placed in the cubic 
volume with $L = 4\;\mu \mathrm{m}$. The calculations have been done for the exact 
Stark-tuned F$\ddot{\mathrm{o}}$rster resonance $37P_{3/2} + 37P_{3/2} \to 37S_{1/2} + 38S_{1/2}$ with zero energy 
defect.}
\end{figure}
\section{Discussion}

We now briefly discuss a possible experimental implementation of the method. 
Micrometer-sized dipole traps can be used to store several atoms and 
to control their positions. By tuning the loading parameters, one can achieve 
control on the average number of the loaded atoms, which can be limited to 
$\overline{N} \approx 1 - 10$. The typical lifetimes of these traps can easily 
reach hundreds of microseconds, and an effective atomic confinement of few 
hundred nanometers can be achieved in all dimensions~\cite{Loading}. Therefore a high 
fidelity of the single-atom excitation should be expected in microscopic 
dipole traps loaded with small number of atoms.

The intense laser field of the dipole trap induces position-dependent light shifts of the atomic energy levels and
can also photoionize Rydberg atoms. The effect of light shifts can be suppressed by using the trapping light with a "magic wavelength",
which matches light shifts of the ground and Rydberg states~\cite{Safronova}.
The photoionization can substantially reduce effective lifetimes of Rydberg atoms~\cite{Shaffer} and this effect
cannot be suppressed with a "magic wavelength". We therefore suggest to avoid both light shifts and photoionization
by temporarily switching the dipole trap off, provided the atom temperatures are sufficiently low ($<50\,\mathrm{\mu K}$) to make the subsequent recapture possible.

To conclude, we have proposed a method for the 
deterministic single-atom excitation to a Rydberg state in mesoscopic 
ensembles of interacting atoms. Chirped laser excitation and off-resonant STIRAP in the full blockade regime have been shown to be insensitive to the 
number of interacting atoms. This  method is well suited to prepare a 
collective excited state in a small dipole-blockaded sample of atoms loaded 
in micrometer sized dipole traps with high fidelity. It could further 
be used for selective single-atom loading of optical lattices and dipole 
trap arrays, which are initially loaded with an unknown number of atoms. 

Our method opens the way to the implementation of scalable quantum registers and 
single-photon sources for quantum information processing with neutral atoms.

\section{Acknowledgements}

We thank M.~Saffman for helpful discussions. 
This work was supported by the RFBR (Grant Nos. 10-02-00133, 10-02-92624), 
by the Russian Academy of Sciences, by the Presidential Grants No. 
MK-6386.2010.2 and MK-3727.2011.2, and by the Dynasty Foundation. S.~B. and 
C.~M. acknowledge support from EPSRC grant EP/F031130/1.

\end{document}